\documentstyle[11pt]{article}
\begin{document}
\begin{titlepage}
\centerline{\large\bf Does Weak CP Phase Originate from a
               Certain Geometry ?}
\vspace{1cm}

\centerline{ Yong Liu }
\vspace{0.5cm}
\centerline{Lab of Numerical Study for Heliospheric Physics (LHP)}
\centerline{Chinese Academy of Sciences}
\centerline{P. O. Box 8701, Beijing 100080, P.R.China}
\vspace{0.2cm}
\centerline{E-mail: yongliu@ns.lhp.ac.cn}
\vspace{2cm}

\centerline{\bf Abstract}
\vspace{0.6cm}

We further investigate the probability that, weak CP phase originates in
a certain geometry. We find that our postulation on weak CP Phase gives
strict constraints on angle $\gamma$ in unitarity triangle {\bf DB} and
the element $V_{td}$ in Cabibbo-Kobayashi-Maskawa (CKM) matrix. The
predicted $|V_{td}|$ is about 0.0086 and $\gamma$ about $75.3^0$ with
the very narrow window respectively.
These two parameters can be used to test our postulation precisely in near
future.
\\\\
PACS number(s): 11.30.Er, 12.10.Ck, 13.25.+m

\end{titlepage}

\centerline{\large\bf Does Weak CP Phase Originate from a
               Certain Geometry ?}
\vspace{1cm}

As is well known, CP violation is one of the most important problem
in particle physics [1-6].
People have been concerning it for more than thirty
years. However, we still know little about it. With the runing of the
$B-$factories, we expect to know more about it in near future.

According to the standard CKM mechanism, CP violation
originates from a phase
presenting in the three by three quark mixing matrix [7-8].
Last winter, we have found that, the weak CP phase and the other
three mixing angles in CKM matrix satisfy a certain geometry relation
[9-10], so we postulated that, weak CP phase is a geometry phase.
Although it is a ad hoc postulation, all the conclusions extracted from
it are consistent with the present experimental results.

Our postulation in the standard parametrization can be expressed as [10]
\begin{equation}
\sin\delta_{13}=\frac{ (1+s_{12}+s_{23}+s_{13})
                       \sqrt{1-s_{12}^2-s_{23}^2-s_{13}^2+
                       2 s_{12} s_{23} s_{13}} }{(1+
                       s_{12}) (1+s_{23}) (1+s_{13})}
\end{equation}
where $s_{ij}$ and $\delta_{13}$ are the parameters in the standard
parametrization [11-12]
\begin{equation}
V_{KM}= \left (
\begin{array}{ccc}
   c_{12} c_{13} & s_{12} c_{13}& s_{13} e^{-i \delta_{13}} \\
   -s_{12} c_{23}-c_{12} s_{23} s_{13} e^{i \delta_{13}} &
   c_{12} c_{23}-s_{12} s_{23} s_{13} e^{i \delta_{13}}    &
   s_{23} c_{13}\\
   s_{12} s_{23}-c_{12} c_{23} s_{13} e^{i \delta_{13}}  &
   -c_{12} s_{23}-s_{12} c_{23} s_{13} e^{i \delta_{13}} &
   c_{23} c_{13}
\end{array}
\right )
\end{equation}
with $c_{ij}=\cos\theta_{ij}$ and $s_{ij}=\sin\theta_{ij}$ for the
"generation" labels $i,j=1,2,3$.
Here, the real
angles $\theta_{12}, \theta_{23}$ and $\theta_{13}$ can all be made to
lie in the first quadrant. The phase $\delta_{13}$ lies in the range
$0<\delta_{13}<2 \pi$. In following, we will make the three angles
$\theta_{ij}$ lie in the first quadrant.

The central purpose of this work is to further
investigate our postulation carefully. Firstly, we investigate
the whole CKM matrix, secondly, we investigate the three angles
in unitarity triangle ${\bf DB}$, both with our postulation
being used. In fact, if we notice that, the moduli of the elements
in CKM matrix are squared moduli invariants, and the three angles
in unitarity triangle are composed of squared and quartic invariants [13],
we can see that is meaningful to do so.

\vspace{0.5cm}
\centerline{\bf 1. About the Moduli of the Elements in CKM Matrix}

We investigate the whole CKM matrix firstly, we hope to extract some
useful information from our postulation. The programme is\\
$a$. For each group of given $V_{ud},\; V_{ub}$ and
$V_{tb}$, solve $s_{12},\;s_{23},\;s_{13}$ from the following equation
\begin{equation}
V_{ud}=c_{12}\; c_{13}\;\;\;\;\;\;
V_{ub}=s_{13} \;\;\;\;\;\;
V_{tb}=c_{23}\; c_{13}.
\end{equation}
$b$. Substituting Eq.(1) into CKM matrix Eq.(2). Then, solve the
moduli of all the elements.\\
$c$. Let $V_{ud},\; V_{ub}$ and $V_{tb}$ vary in certain ranges.
Repeat the steps $a$ and $b$.

When we let $V_{ud},\; V_{ub}$ and $V_{tb}$ vary in the ranges [11]
\begin{equation}
0.9745 \leq V_{ud} \leq 0.9760 \;\;\;\;\;\;
0.0018 \leq V_{ub} \leq 0.0045 \;\;\;\;\;\;
0.9991 \leq V_{tb} \leq 0.9993
\end{equation}

\begin{equation}
0.9745 \leq V_{ud} \leq 0.9760 \;\;\;\;\;\;
0.00045 \leq V_{ub} \leq 0.00585 \;\;\;\;\;\;
0.9991 \leq V_{tb} \leq 0.9993
\end{equation}
and
\begin{equation}
0.97375 \leq V_{ud} \leq 0.97675 \;\;\;
0.00045 \leq V_{ub} \leq 0.00585 \;\;\;
0.99895 \leq V_{tb} \leq 0.99955 \;\;\;
\end{equation}

we get the magnitudes of the elements of the complete matrix
as
\begin{equation}
\left (
\begin{array}{ccc}
0.9745 \sim 0.9760    &0.2182 \sim 0.2244  & 0.0018 \sim  0.0045\\
0.2180 \sim 0.2242    &0.9736 \sim 0.9752  & 0.0371 \sim 0.0424 \\
0.0079 \sim 0.0093    &0.0365 \sim 0.0415  & 0.9991 \sim 0.9993
\end{array}
\right )
\end{equation}

\begin{equation}
\left (
\begin{array}{ccc}
0.9745 \sim 0.9760    &0.2182 \sim 0.2244  & 0.00045 \sim  0.00585\\
0.2180 \sim 0.2242    &0.9736 \sim 0.9757  & 0.0296 \sim 0.0424 \\
0.0079 \sim 0.0094    & 0.0364 \sim 0.0415 & 0.9991 \sim 0.9993
\end{array}
\right )
\end{equation}
and
\begin{equation}
\left (
\begin{array}{ccc}
0.97375 \sim 0.97675    &0.2143\sim 0.2276  & 0.00045 \sim  0.00585\\
0.2142 \sim 0.2275    &0.9727 \sim 0.9762  & 0.0296 \sim 0.0458 \\
0.0072\sim 0.0102    &0.0337 \sim 0.0448  & 0.99895 \sim 0.99955
\end{array}
\right )
\end{equation}
respectively.

Compare with that given by [11]
\begin{equation}
\left (
\begin{array}{ccc}
0.9745 \sim 0.9760  & 0.217 \sim 0.224  & 0.0018 \sim  0.0045\\
0.217 \sim 0.224   & 0.9737 \sim 0.9753  & 0.036 \sim 0.042 \\
0.004 \sim 0.013   & 0.035 \sim 0.042 & 0.9991 \sim 0.9993
\end{array}
\right )
\end{equation}
we find that, the predicted results
are well in agreement with that given by data book. In the meantime,
$|V_{td}|$ is not sensitive to the variations of the inputs. It lies
in a very narrow window with the central value about ($\sim 0.0086$),
even if we take a little more large error ranges
for the inputs. 

On the other hand, the relevant result extracted from the
experiment on $B_d^0-\overline{B_d^0}$ mixing is [11]
\begin{equation}
|V_{tb}^* \cdot V_{td}|=0.0084 \pm 0.0018.
\end{equation}
we find that, the prediction about $|V_{td}|$
based on our postulation, not only
coincide with the experimental result very well, but also gives
a more strict constraint.

\vspace{0.5cm}
\centerline{\bf 2. About the Three Angles of the Unitarity Triangle}

The three angles $\alpha,\; \beta$ and $\gamma$
in the unitarity triangle ${\bf DB}$ defined as [6]
\begin{equation}
\alpha \equiv arg(-\frac{V_{td} V_{tb}^*}{V_{ud} V_{ub}^*})\;\;\;\;\;\;
\beta \equiv arg(-\frac{V_{cd} V_{cb}^*}{V_{td} V_{tb}^*}) \;\;\;\;\;\;
\gamma \equiv arg(-\frac{V_{ud} V_{ub}^*}{V_{cd} V_{cb}^*})
\end{equation}

If a small change being made, it is easy to see that, the definited
angles are composed of squared and quartic invariants. For example,
$$
\alpha \equiv arg(-\frac{V_{td} V_{tb}^*}{V_{ud} V_{ub}^*})
=arg(-\frac{V_{td} V_{ud}^* V_{tb}^* V_{ub} }
{V_{ud} V_{ud}^* V_{ub} V_{ub}^*})
$$
the numerator in the definition is a quartic invariant
and the denominator is a product of two squared invariants.

Now, we investigate the three angles. The programme is similar to
that in section {\bf 1}.\\
$a$. For each group of given $V_{ud},\; V_{ub}$ and
$V_{tb}$, solve $s_{12},\;s_{23},\;s_{13}$ from Eq.(3).\\
$b$. Substituting Eq.(1) into CKM matrix Eq.(2). Then, solve
all of the elements with the results of $a$ being used.\\
$c$. Solve $\alpha,\; \beta$ and $\gamma$ according to the
definition Eq.(12).\\
$d$. Let $V_{ud},\; V_{ub}$ and $V_{tb}$ vary in certain ranges.
Repeat the steps $a$, $b$ and $c$.

We still let $V_{ud},\; V_{ub}$ and $V_{tb}$ vary in the ranges
given by Eq.(4), Eq.(5) and Eq.(6). The corresponding results are
\begin{equation}
73.3^0 \leq \alpha \leq 94.4^0 \;\;\;\;\;\;
10.6^0 \leq \beta \leq 31.3^0 \;\;\;\;\;\;
74.9^0 \leq \gamma \leq 75.6^0
\end{equation}

\begin{equation}
70.3^0 \leq \alpha \leq 98.9^0 \;\;\;\;\;\;
6.1^0 \leq \beta \leq 34.2^0\;\;\;\;\;\;
74.8^0 \leq \gamma \leq 75.6^0
\end{equation}
and
\begin{equation}
66.9^0 \leq \alpha \leq 99.8^0 \;\;\;
5.6^0 \leq \beta \leq 34.2^0 \;\;\;
74.5^0 \leq \gamma \leq 76.0^0 \;\;\;
\end{equation}
respectively.

The recent analysis by Buras gives [14]
\begin{equation}
35^0 \leq \alpha \leq 115^0 \;\;\;\;\;\;
11^0 \leq \beta \leq 27^0\;\;\;\;\;\;
41^0 \leq \gamma \leq 134^0
\end{equation}
or more strictly
\begin{equation}
70^0 \leq \alpha \leq 93^0 \;\;\;
19^0 \leq \beta \leq 22^0 \;\;\;
65^0 \leq \gamma \leq 90^0 \;\;\;
\end{equation}

It is easy to find, similar to that on $V_{td}$ in above,
starting from our postulation, we obtain a more strict
constraint on $\gamma$. We predict a very narrow window for $\gamma$
with the central value about
($ \sim 75.3^0 $).  Furthermore, all the predictions about $\alpha,\;
\beta$ and $\gamma$ coincide with the relevant analysis [14].

\vspace{0.5cm}
\centerline{\bf 3. Conclusions and Discussions}

In conclusion, we have further investigated the postulation that, weak CP
phase originates in a certain geometry. Based on this postulation, we
obtian two strict constraints on the magnitude of CKM matrix
element $V_{td}$ and angle $\gamma$ respectively. These can be put to
the more precisely test on our postulation in $B-$factory in near future.

The significance of Eq.(1) is evident. If it can be further verified
in the future, we can at least remove a uncertainty coming from the
weak interaction. Through the study on heavy flavors, with Eq.(1) being
considered, we can then extract more imformations on strong interaction.
For instance, it becomes possible to extract the CP phase related to
the part of strong interaction of final states. The further work
along this direction is under way.

\end{document}